\documentclass[aps, amsmath,amssymb,prc,twocolumn,superscriptaddress]{revtex4-1}

\usepackage{graphicx,color}
\usepackage{dcolumn}
\usepackage{bm}

\usepackage{import}

\begin{document}
  \newcommand {\nc} {\newcommand}
  \nc {\beq} {\begin{eqnarray}}
  \nc {\eeq} {\nonumber \end{eqnarray}}
  \nc {\eeqn}[1] {\label {#1} \end{eqnarray}}
  \nc {\eol} {\nonumber \\}
  \nc {\eoln}[1] {\label {#1} \\}
  \nc {\ve} [1] {\mbox{\boldmath $#1$}}
  \nc {\ves} [1] {\mbox{\boldmath ${\scriptstyle #1}$}}
  \nc {\mrm} [1] {\mathrm{#1}}
  \nc {\half} {\mbox{$\frac{1}{2}$}}
  \nc {\thal} {\mbox{$\frac{3}{2}$}}
  \nc {\fial} {\mbox{$\frac{5}{2}$}}
  \nc {\la} {\mbox{$\langle$}}
  \nc {\ra} {\mbox{$\rangle$}}
  \nc {\etal} {\emph{et al.}}
  \nc {\eq} [1] {(\ref{#1})}
  \nc {\Eq} [1] {Eq.~(\ref{#1})}
  \nc {\Ref} [1] {Ref.~\cite{#1}}
  \nc {\Refc} [2] {Refs.~\cite[#1]{#2}}
  \nc {\Sec} [1] {Sec.~\ref{#1}}
  \nc {\chap} [1] {Chapter~\ref{#1}}
  \nc {\anx} [1] {Appendix~\ref{#1}}
  \nc {\tbl} [1] {Table~\ref{#1}}
  \nc {\Fig} [1] {Fig.~\ref{#1}}
  \nc {\ex} [1] {$^{#1}$}
  \nc {\Sch} {Schr\"odinger }
  \nc {\flim} [2] {\mathop{\longrightarrow}\limits_{{#1}\rightarrow{#2}}}
  \nc {\textdegr}{$^{\circ}$}
  \nc {\inred} [1]{\textcolor{red}{#1}}
  \nc {\inblue} [1]{\textcolor{blue}{#1}}
  \nc {\IR} [1]{\textcolor{red}{#1}}
  \nc {\IB} [1]{\textcolor{blue}{#1}}
  \nc{\pderiv}[2]{\cfrac{\partial #1}{\partial #2}}
  \nc{\deriv}[2]{\cfrac{d#1}{d#2}}
  
  \nc {\ai}{\emph{ab~initio}}


\title{Systematic analysis of the peripherality of the $^{10}$Be$(d,p)$$^{11}$Be transfer reaction and extraction of the asymptotic normalization coefficient of $^{11}$Be bound states.}

\author{J. Yang} 
 \email{jiecyang@ulb.ac.be}
\affiliation{Physique Nucl\' eaire et Physique Quantique (CP 229), Universit\'e libre de Bruxelles (ULB), B-1050 Brussels}
\affiliation{Afdeling Kern-en Stralingsfysica, Celestijnenlaan 200d-bus 2418, 3001 Leuven, Belgium}
\author{P. Capel}
\email{pcapel@uni-mainz.de}
 \affiliation{Institut f\"ur Kernphysik, Johannes Gutenberg-Universit\"at Mainz, 55099 Mainz, Germany}
\affiliation{Physique Nucl\' eaire et Physique Quantique (CP 229), Universit\'e libre de Bruxelles (ULB), B-1050 Brussels}

\date{\today}

\begin{abstract}
We reanalyze the experiment of Schmitt \etal\ on the $^{10}$Be$(d,p)^{11}$Be transfer reaction [Phys. Rev. Lett. \textbf{108}, 192701 (2012)] by exploring the beam-energy and angular ranges at which the reaction is strictly peripheral.
We consider the adiabatic distorted wave approximation (ADWA) to model the reaction and use a Halo-EFT description of $^{11}$Be to systematically explore the sensitivity of our calculations to the short-range physics of the $^{10}$Be-$n$ wave function.
We find that by selecting the data at low beam energy and forward scattering angle the calculated cross sections scale nearly perfectly with the asymptotic normalization coefficient (ANC) of the $^{11}$Be bound states.
Following these results, a comparison of our calculations with the experimental data gives a value of $C_{1s1/2}=0.785\pm0.03$~fm$^{-1/2}$ for the $\half^+$ ground-state ANC and $C_{0p1/2}=0.135\pm0.005$~fm$^{-1/2}$ for the $\half^-$ excited-state, which are in perfect agreement with the \ai\ calculations of Calci \etal, who obtain $C^{\textit{ab initio}}_{1/2^+}=0.786$~fm$^{-1/2}$ and $C^{\textit{ab initio}}_{1/2^-}=0.129$~fm$^{-1/2}$ [Phys. Rev. Lett. \textbf{117}, 242501 (2016)].

\begin{description}
\item[Keywords]
Transfer reaction, halo nuclei, $^{11}$Be, ADWA, ANC.
\item[PACS numbers] 
21.10.Gv, 21.10.Jx, 25.45.Hi, 25.60.Je
\end{description}
\end{abstract}

\pacs{Valid PACS appear here}

\maketitle

\section{\label{intro}Introduction}

Halo nuclei \cite{halo} constitute a unique class of exotic systems, which are mainly found in the neutron-rich region of the nuclear chart.
The halo is a threshold effect observed close to the neutron dripline, in which one or two neutrons are loosely bound to the core of the nucleus.
Due to this loose binding, these valence neutrons can tunnel far away into the classically forbidden region and exhibit a high probability of presence at a large distance from the other nucleons.
They hence form a sort of diffuse halo around a compact core \cite{HJ87}, which significantly increases the matter radius of these nuclei.

Since their discovery in the mid-80s, halo nuclei have been the subject of many studies in both the nuclear-structure and nuclear-reaction communities.
In the former because of the challenge these diffuse nuclei pose to usual nuclear-structure models, like the shell model.
In the latter because, due to their short lifetime, they are mostly studied through reactions.

Experimentally, the upgrade of rare isotope beam facilities worldwide provides us with many ways to explore these halo systems.
Transfer reaction \cite{AN03,thompson,Joh14,euroschool,Pot17,Wim18} has been an important tool to infer information about these systems for decades.
In this reaction, one or several nucleons are transferred between the projectile and target.
Since those nucleons populate the valence states of the nucleus, transfer is useful in the analysis of the single-particle structure of nuclei \cite{AN03,thompson,Jon10,schmitt,schmittc,Wim18}.
It is therefore particularly well suited to study halo nuclei \cite{euroschool,schmitt,schmittc,MMT14,belyaeva}.

To extract valuable nuclear-structure information from experimental data, a precise model of the reaction is required.
Deuteron-induced reactions, like the one on which this work is focused, are usually described within a three-body model: a proton $p$, a neutron $n$ and the nucleus upon which the transfer takes place.
Many such models have been developed \cite{AN03,thompson,Joh14,euroschool,Pot17}.
The Distorted Wave Born Approximation (DWBA) \cite{dwba} is one of the most used methods to analyze experimental data and extract spectroscopic information about nuclei.
However, this method does not properly account for dynamical effects, such as the breakup of the deuteron, therefore alternative formulations have been suggested.
Johnson and Soper have introduced the adiabatic distorted wave approximation (ADWA), which, without losing the relative simplicity of the DWBA method, includes a zero-range adiabatic treatment of the deuteron-breakup channel (ZR-ADWA) \cite{soper}.
Johnson and Tandy have then extended this seminal work to a finite-range version of the ADWA method (FR-ADWA) \cite{tandy}.
For a more accurate inclusion of the deuteron dynamics in the reaction model, the solution of the continuum-discretized coupled-channel approach (CDCC) \cite{Aus87} can be used.
In that approach, the projectile-target wave function is expanded upon all the states of the deuteron, including its continuum, which leads to the resolution of a set of coupled equations.
More recently, numerical techniques have become available to solve the Faddeev-Alt,
Grassberger, and Sandhas (FAGS) equations \cite{Fad61,AGS67}, which corresponds to the most accurate framework to describe transfer reactions induced by deuteron within a three-body model \cite{Del09}.


At the Oak Ridge National Laboratory a transfer experiment has been performed by Schmitt \etal\ to study the structure of $^{11}$Be \cite{schmitt,schmittc}.
This nucleus is the archetypical one-neutron halo nucleus and, as such, exhibits a strong $^{10}$Be-$n$ structure.
In this Oak-Ridge experiment a neutron is transferred from a deuteron to $^{10}$Be to form $^{11}$Be: $^{10}$Be$(d,p)^{11}$Be.
The two bound states of $^{11}$Be have been populated: its $\half^+$ ground state and  $\half^-$ excited state.
Transfer to the $\fial^+$ resonance above the one-neutron threshold has also been measured.
The experiment has been performed in inverse kinematics with an ultra-pure $^{10}$Be beam impinging on a CD$_2$ target at beam energies 107, 90, 75, and 60~MeV, which correspond, in direct kinematics, to, respectively, $E_d=21.4$, 18, 15, and 12~MeV in the laboratory restframe \cite{schmittc}.

The main goal of the present work is to reanalyze this Oak-Ridge experiment with a special focus on the sensitivity of the calculations to the $^{10}$Be-$n$ wave function in the $^{11}$Be bound states.
In particular, we look for the best experimental conditions in which the reaction is strictly peripheral, i.e. for which only the tail of the $^{10}$Be-$n$ radial wave function affects the theoretical cross sections.
Since this tail has a universal behavior \cite{Tim14}, but for its normalization, the comparison with the data in these peripheral conditions should enable us to extract this asymptotic normalization constant (ANC) in a model-independent way \cite{BDB77,Gag99,MN05,pang,INY07,MMT14,belyaeva}.

To reach this goal, we couple a Halo-EFT description of $^{11}$Be \cite{BHK02,HJP17} to the ADWA model of reaction.
Thanks to the natural separation of scales in EFT, this provides us with a very systematic way of studying the sensitivity of the cross section to the short-range physics of the overlap wave function.
Albeit similar in spirit with Refs.~\cite{MN05,pang,MMT14,belyaeva}, this analysis will enable us to determine the exact conditions of peripherality of the reaction, and hence extract a reliable estimate of the ANC of the bound states of $^{11}$Be.

Recently an \ai\ calculation of $^{11}$Be has been performed by Calci \etal\ within the framework of the no-core shell model with continuum (NCSMC) \cite{abinitio}.
These calculations provide a fully microscopic prediction of its ANC, to which we will be able to confront our values inferred from the data of Schmitt \etal\ \cite{schmitt,schmittc}.

This paper is structured as follows: In \Sec{theory}, we briefly present the three-body model of the reaction and the ADWA, which we use to compute the transfer cross sections.
In \Sec{potentials}, we introduce the numerical inputs and the descriptions of $^{11}$Be we consider in this study.
Finally we present the results of our calculations and discuss them in \Sec{results}.
Our conclusions are drawn in \Sec{conclusion}.

\section{\label{theory}Theoretical Framework}

\begin{figure}[!h]
\centering
\includegraphics[width=\linewidth]{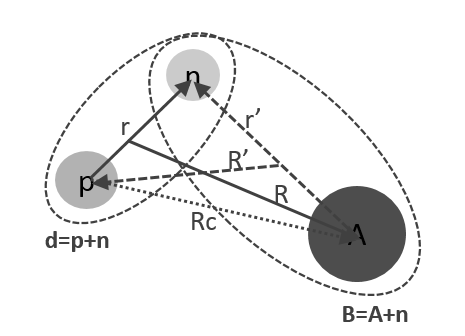}
\caption{Illustration of the three-body system with associated coordinates.}
\label{f1}      
\end{figure}

We consider the stripping reaction $A(d,p)B$ in which a neutron is transferred to a nucleus $A$ ($^{10}$Be) to form nucleus $B$ ($^{11}$Be).
In a simple physical picture, this transfer reaction can be viewed as a process in which the neutron $n$ from the incident deuteron $d$ populates an unoccupied state in the target nucleus $A$, producing the composite nucleus described as a two-cluster structure $B=A+n$.
To model this reaction, we adopt the three-body model ($A+n+p$) illustrated in Fig.~\ref{f1}.

In its post form, the transition matrix elements for the reaction reads \cite{Joh14,euroschool,thompson}
\begin{equation}
T_{\rm post}(pB,dA)=\langle \chi_{pB}^{(-)} \varphi_{An} | V_{pn}+U_{pA}-U_{pB} | \psi_{dA}^{(+)}\rangle,
\label{e1}
\end{equation}
where $V_{pn}$ is the potential that simulates the interaction that binds the proton and the neutron into the deuteron and $U_{pA}$ and $U_{pB}$ are optical potentials that simulates the interaction between the proton and the clusters $A$ and $B$, respectively.
The wave function $\psi_{dA}^{(+)}$ describes the three-body system with the condition that the proton and neutron are initially bound into a deuteron that is impinging on $A$.
At the ADWA, it is approximated by
\begin{equation}
\psi_{dA}^{(+)}(\bm{r},\bm{R})\simeq \chi_{dA}^{(+)}(\bm{R}) \varphi_{pn}(\bm{r}),
\label{e3}
\end{equation}
where $\varphi_{pn}$ is the deuteron bound state computed from $V_{pn}$ and $\chi_{dA}^{(+)}$ is the distorted wave describing the scattering of $d$ by $A$.
Following the Johnson and Tandy prescription \cite{tandy}, this wave function is obtained from the optical potential $U_{dA}$ built by averaging $A$-$p$ and $A$-$n$ optical potentials over the finite-range deuteron bound state
\begin{equation}
U_{dA}(R)= \frac{\langle \varphi_{pn} | V_{pn}(U_{pA}+U_{nA}) | \varphi_{pn}\rangle}{\langle \varphi_{pn} | V_{pn} | \varphi_{pn}\rangle}.
\label{e3}
\end{equation}

The distorted wave $\chi_{pB}^{(-)}$ appearing in \Eq{e1} describes the scattering of $p$ by the cluster $B$ in the outgoing channel of the reaction; it is obtained using the optical potential $U_{pB}$.
The wave function $\varphi_{An}$ describes the state of the nucleus $B$ formed in the transfer.
In this three-body model, it is obtained at the single-particle approximation, in which $B$ is seen as a two-cluster structure, in which a neutron is bound to the core $A$ assumed to be structureless.
The $A$-$n$ interaction is described by a phenomenological potential $V_{An}$.
Following Refs.~\cite{CPH18,varna17}, we use a Halo-EFT description of $^{11}$Be (see \Sec{Be11}) \cite{BHK02,HJP17}.
Within this description, the $B$ bound state is characterized by the quantum numbers $n_{r'}lj$, where $n_{r'}$ is the number of nodes in the radial wave function, $l$ is the orbital angular momentum and $j$ is obtained from the coupling of $l$ with the spin of the neutron.

The reduced radial wave function has the following asymptotic behavior
\begin{equation}
u_{n_{r'}lj}(r') \flim{r'}{\infty} b_{n_{r'}lj}\ i\,\kappa_{n_{r'}lj}\,r'\ h_l(i\kappa_{n_{r'}lj} r'),
\label{e4}
\end{equation}
where $h_l$ is a spherical Hankel function function and $\kappa_{n_{r'}lj}=\sqrt{2 \mu_{An} |E_{n_{r'}lj}|}/\hbar$, with $|E_{n_{r'}lj}|$ the binding energy of the neutron to the core $A$ and $\mu_{An}$ their reduced mass.
The parameter $b_{n_{r'}lj}$ is the single-particle ANC (SPANC) defining the strength of the exponential tail of the $A $-$n$ bound-state wave function.
This SPANC will vary with the geometry of the potential used to simulate the $A$-$n$ interaction \cite{Tim14,MMT14,belyaeva,CN06}.
We will use this property in \Sec{results} to assess the sensitivity of the transfer cross section to the ANC.

Being universal, the asymptotic behavior \eq{e4} exists also in the actual structure of the nuclei \cite{Tim14} and hence should be reproduced in \ai\ models, like the NCSMC calculation of Calci \etal\ \cite{abinitio}.
However, the true ANC will differ from the SPANC obtained in the phenomenological two-body description of $B$ due to the coupling with the other possible configurations \cite{Tim14}.
In the present piece of work, we study how to relate the two and if there are experimental conditions which enable a safe extraction of the ANC for the $^{11}$Be bound states from the Oak-Ridge experiment \cite{schmitt,schmittc}.

The theoretical differential cross section expressed as a function of the relative direction $\Omega=(\theta,\phi)$ between $p$ and $B$ in the outgoing channel $d\sigma_{\rm th}/d\Omega$ is obtained from the square modulus of the transition matrix elements \eq{e1}.
All transfer calculations are performed with the code FRESCO \cite{fresco}.
In the next section, we provide all the details about our choices of the potentials used in this work.

\section{\label{potentials}Two-body potentials}

\subsection{\label{Be11}Description of $^{11}$Be}

As mentioned in the previous sections, $^{11}$Be is the archetype of a one-neutron halo nucleus.
It can thus be modeled as a neutron loosely bound to a $^{10}$Be core.
With the assumption that the $^{10}$Be core is in its ground state ($0^{+}$), the $\half^{+}$ ground state (g.s.) of $^{11}$Be can be described by a $^{10}{\rm Be}(0^{+})\otimes1s_{1/2}$ configuration, and the $\half^{-}$ excited state (e.s.) by a $^{10}{\rm Be}(0^{+})\otimes0p_{1/2}$ configuration.
In this study, we use a Halo-EFT description of this nucleus at the leading order of the expansion in each of these partial waves \cite{BHK02,HJP17}.

Halo EFT provides a systematic treatment of halo nuclei, which exhibit a clear separation of scales: the core of the nucleus ($^{10}$Be in the present case) is tightly bound and hence compact, whereas the halo neutron is loosely bound and consequently has a very extended wave function.
The parameter $R_{\rm core}/R_{\rm halo}$, where $R_{\rm core}$ ($R_{\rm halo}$) is the size of the core (halo) of the nucleus, is thus small (about 0.4 for $^{11}$Be).
Halo EFT exploits this separation of scales and considers the core and halo neutron as its degrees of freedom.
Within Halo EFT, the quantum-mechanical amplitudes are expanded into powers of that parameter (see Ref.~\cite{HJP17} for a recent review).
This effective theory will break down if the process it describes probes distances smaller than $R_{\rm core}$, or if they lead to the excitation of the core.

Halo EFT is expressed through a Lagrangian that includes all operators up to a given order in this expansion.
The interactions that appear in this Lagrangian are thus considered at the limit $R_{\rm core}/R_{\rm halo}\rightarrow0$ and are described by zero-range potentials and their derivatives.
The coefficients of these potentials---the low-energy constants of the theory---are free parameters, which are adjusted to reproduce experimental data or outputs of \emph{ab initio} calculations \cite{CPH18}.
In the present work, we consider the development at the lowest order using just one contact term, and hence one low-energy constant, per partial wave to simply reproduce the one-neutron separation energy of each bound state of $^{11}$Be populated through the transfer reactions measured by Scmitt \etal\ \cite{schmitt,schmittc}.
We neglect the possible derivatives of the interaction as well as the higher-oder terms \cite{BHK02,HJP17}.

To render the interactions numerically tractable, we follow what has been done to describe the nucleon-nucleon interaction in EFT \cite{Kievsky:2016kzb} and regulate them with a Gaussian, whose range can be varied \cite{CPH18,varna17}
\begin{equation}
V_{An}(r')=-V_0 \ e^{-\frac{r'^2}{2r_0^2}}.
\label{e6}
\end{equation}
This form of the neutron-core potential enables us to easily evaluate the sensitivity of the reaction to the short-range physics, which is believed to take place at distances shorter than the radial range $\sqrt{2}\,r_0$ of these Gaussians.
Our goal being to find the experimental conditions under which the reaction is purely peripheral, halo EFT provides us with a simple and elegant tool to generate, using different values of the Gaussian width $r_0$, wave functions for the bound states of $^{11}$Be that exhibit significantly different radial behaviors.
For the reaction to be peripheral, it needs to be sensitive only to the tail of the radial wave function \eq{e4}.
One simple way to find that out is to check that its cross section is proportional to the square of the bound state SPANC $|b_{n_{r'}lj}|^2$, using different $A$-$n$ potentials that generate single-particle wave functions with different SPANCs, as was already done in Refs.~\cite{MN05,CN06,pang,MMT14,belyaeva}.
However, we must also be sure that the reaction is \emph{not} sensitive to the internal part of the wave function.
For this, the different wave functions must not only have different SPANCs, but should also exhibit very different radial behavior inside the nucleus.

The Gaussian potential \eq{e6}, enables us to realize that in a simple way.
We consider nine such Gaussian potentials with different widths $r_{0}$ ranging from 0.4~fm to 2.0~fm.
The lower end of that range is unphysically small, but it enables us to generate both very small SPANCs and significant changes in the internal part of the wave function.
The upper end is chosen so as to avoid distortion in the long-range physics of $^{11}$Be \cite{CPH18}.

For each width the depth $V_0$ in the $s_{1/2}$ partial wave is adjusted to reproduce the neutron binding energy: $|E_{1s1/2}|=0.502$~MeV for the g.s. \cite{KKP12}.
We do the same in the $p_{1/2}$ partial wave to describe the $\half^-$ bound excited state of $^{11}$Be, fitting the depth of the central term $V_0$ to obtain $E_{0p1/2}=-0.182$~MeV \cite{KKP12}.
These parameters are listed in \tbl{t1} with the corresponding SPANCs $b_{1s1/2}$ and $b_{0p1/2}$.
This way of doing enables us to generate a very broad range of SPANCs for both the ground and excited bound states of $^{11}$Be.

\begin{table}
\caption{\label{t1}
Parameters of the Gaussian $^{10}$Be-$n$ potentials [See Eq.(\ref{e6})] adjusted to reproduce the g.s.\ and e.s.\ of $^{11}$Be.
The SPANC $b_{n_{r'}lj}$ obtained for each case is provided as well.}
\begin{ruledtabular}
\begin{tabular}{c|cc|cc}
$r_0$ & $V_0$ (g.s.) & $b_{1s1/2}$ & $V_0$ (e.s.) & $b_{0p1/2}$ \\
   (fm) & (MeV) & (fm$^{-1/2}$) & (MeV) & (fm$^{-1/2}$) \\
\hline
0.4 & 1314.6 & 0.601 & 869.4 & 0.068 \\
0.6 & 592.3  & 0.632 & 387.3 & 0.085 \\
0.8 & 337.8  & 0.664 & 218.4 & 0.100 \\
1.0 & 219.2  & 0.697 & 140.2 & 0.114 \\
1.2 & 154.4  & 0.732 & 97.7  & 0.127 \\
1.4 & 115.1  & 0.769 & 72.1  & 0.140 \\
1.6 & 89.3   & 0.807 & 55.4  & 0.152 \\
1.8 & 71.6   & 0.846 & 44.0  & 0.165 \\
2.0 & 58.8   & 0.888 & 35.8  & 0.177 \\
\end{tabular}
\end{ruledtabular}
\end{table}

\begin{figure}
\centering
\includegraphics[width=\linewidth]{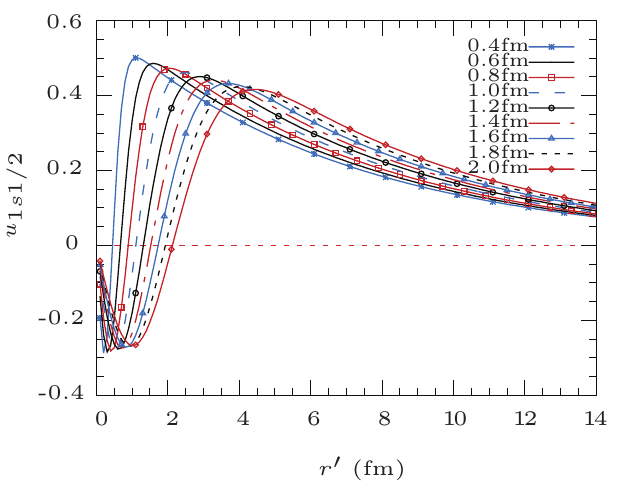}
\caption{Reduced radial wave functions $u_{1s1/2}$ of the $\half^{+}$ g.s.\ of $^{11}$Be obtained with the nine Gaussian potentials of \tbl{t1}.}
\label{f2}
\end{figure}

\begin{figure}
\centering
\includegraphics[width=\linewidth]{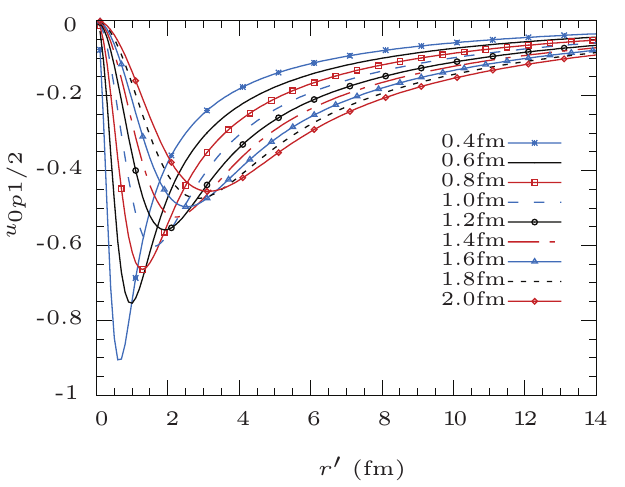}
\caption{Reduced radial wave functions $u_{0p1/2}$ of the $\half^{-}$ e.s.\ of $^{11}$Be obtained with the nine Gaussian potentials of \tbl{t1}.}
\label{f3}
\end{figure}

The corresponding reduced radial wave functions are displayed in Figs.~\ref{f2} and \ref{f3} for the g.s.\ and the e.s., respectively.
As desired for this study, we observe that the nine Gaussian potentials provide radial wave functions significantly different from one another.
The very narrow potentials lead to wave functions that reach their asymptotic behavior \eq{e4} at quite a small radius, viz. $r'\simeq1$~fm, while the broader ones have their internal behavior developing at much larger distances.
The wave function corresponding to $r_0=2.0$~fm being similar to what a usual Woods-Saxon potential produces, i.e. with an asymptotic behaviour reached at $r'\simeq5$~fm (see, e.g., Fig.~7 of \Ref{belyaeva} or Fig.~6(a) of \Ref{CN06}).
These significant changes in both the SPANCs and in the radial behavior in the interior of the nucleus, will help us assessing the sensitivity of our $^{10}$Be$(d,p)$$^{11}$Be transfer calculations to the radial wave function of the $^{11}$Be bound states.
In particular, let us note that these wave functions differ very significantly in the surface part of the nucleus---viz. at $r'\sim 2$--3~fm---to which transfer reactions can be sensitive \cite{Tim14,pang}.
The study of the transfer calculations performed with these very different wave functions will enable us to clearly identify the experimental conditions under which the reaction is purely peripheral.

\subsection{\label{numeric}Other optical potentials}

The nucleon-nucleus optical potentials used to compute the distorted waves used in \Eq{e1} and to build the FR-ADWA $d$-$A$ potential in \Eq{e3} are obtained from the global Chapel-Hill parametrization CH89 \cite{ch89} without including the spin-orbit terms.
This potential is energy dependent and hence needs to be adapted as a function of the deuteron energy $E_d$.
The FR-ADWA potential \eq{e3} is obtained by computing $U_{p^{10}\rm{Be}}$ and $U_{n^{10}\rm{Be}}$ at half the deuteron energy.
For that potential, the numerical integration is performed with the front-end code TWOFNR \cite{twofnr}. 

To test the sensitivity of our calculations to the choice of these optical potentials, we also consider the Koning-Delaroche parametrization \cite{kd}.
The results of these tests are presented in \Sec{sensitivity}.

The Reid soft-core interaction \cite{reid} is used as $V_{pn}$.

\section{\label{results}Results and Discussion}

Following the experimental conditions of Refs.~\cite{schmitt,schmittc}, we perform ADWA calculations of the reaction $^{10}$Be$(d,p)^{11}$Be at energies $E_d=21.4$, 18, 15, and 12~MeV.
We first consider the transfer towards the g.s.\ (\Sec{gs}) and then towards the e.s.\ (\Sec{es}).
In both cases, we study the experimental conditions for which the reaction is peripheral and accordingly extract an ANC for each of these states, which we then compare to the prediction of the \ai\ calculations of Calci \etal\ \cite{abinitio}.

\subsection{\label{gs}Transfer to $^{11}$Be ground state}
\subsubsection{Conditions of peripherality of the reaction}

\begin{figure}
\centering
\includegraphics[width=\linewidth]{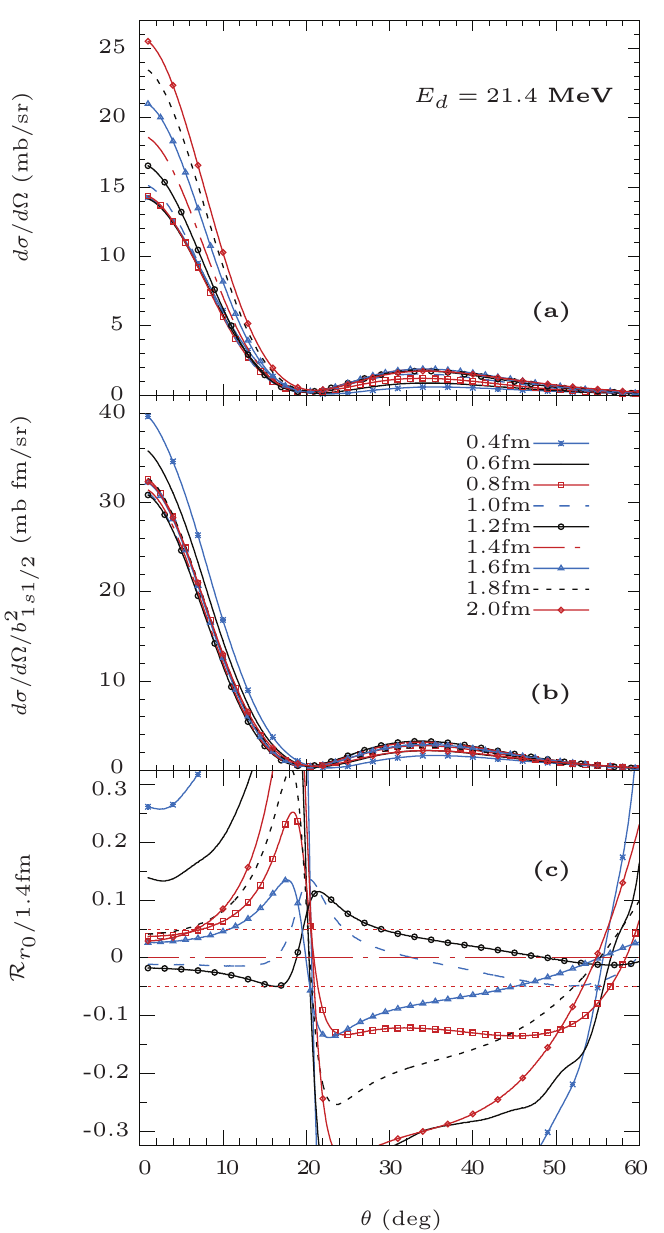}
\caption{Analysis of the differential cross sections of $^{10}$Be$(d,p)^{11}$Be(g.s.) for a deuteron energy $E_d=21.4$~MeV.
The results of the ADWA calculations are presented for every potential of \tbl{t1}.}
\label{f4}
\end{figure}

Figure~\ref{f4}(a) displays the differential cross section $d\sigma_{\rm th}/d\Omega$ for the transfer to the $^{11}$Be g.s.\ computed for the highest experimental deuteron energy $E_d=21.4$~MeV.
The calculations have been performed for the nine $1s_{1/2}$ wave functions shown in \Fig{f2} obtained with the potentials of \tbl{t1}.
As expected, we observe a large variation in the results.
At forward angle, the cross sections seem to scale with the square of the SPANC $b_{1s1/2}$ (see \tbl{t1}), as one would expect if the process were purely peripheral [see \Eq{e1}].
At larger angle, i.e. in the region of the second peak, the ordering of the curves is inverted, showing that in this angular range, the process is more sensitive to the short-range physics of the wave function.
Therefore, selecting data at small scattering angle might enable us to constrain the g.s.\ ANC.

To better estimate the sensitivity of our calculations to the SPANC, we have plotted in \Fig{f4}(b) the transfer cross section divided by $b^2_{1s1/2}$.
Accordingly, the spread in the results is significantly reduced at forward angle, confirming our initial impression of \Fig{f4}(a).
In the region of the second maximum, however, it remains similar to what was observed before scaling.

To precisely determine within which angular range the data should be limited to select a strictly peripheral process, we remove the major angular dependence by considering the following ratio
\begin{equation}
{\cal R}_{r_0/1.4{\rm fm}}(\theta)=\left(\frac{b^{(1.4{\rm fm})}_{n_{r'}lj}}{b^{(r_0)}_{n_{r'}lj}}\right)^2\frac{\frac{d\sigma^{(r_0)}_{\rm th}}{d\Omega}}{\frac{d\sigma^{(1.4{\rm fm})}_{\rm th}}{d\Omega}}-1,
\label{e7}
\end{equation}
where the transfer cross section computed using the $^{10}$Be-$n$ Gaussian potential of width $r_0$ scaled by the square of the SPANC is divided by the result obtained with $r_0=1.4$~fm, which is at the center of the range in $r_0$.
The results are displayed in \Fig{f4}(c).
If one excepts the very narrow potentials ($r_0=0.4$~fm and $r_0=0.6$~fm), we see that all ratios ${\cal R}_{r_0/1.4{\rm fm}}$ fall very close to one another, confirming the peripherality of the reaction when the data are selected at forward angles.
To define an angular range in which the reaction can be considered as peripheral, we consider a maximum of 5\% difference [horizontal dotted lines in \Fig{f4}(c)].
In this case, this happens only at very forward angles, viz. when $\theta<7^\circ$.

\begin{figure}
\centering
\includegraphics[width=\linewidth]{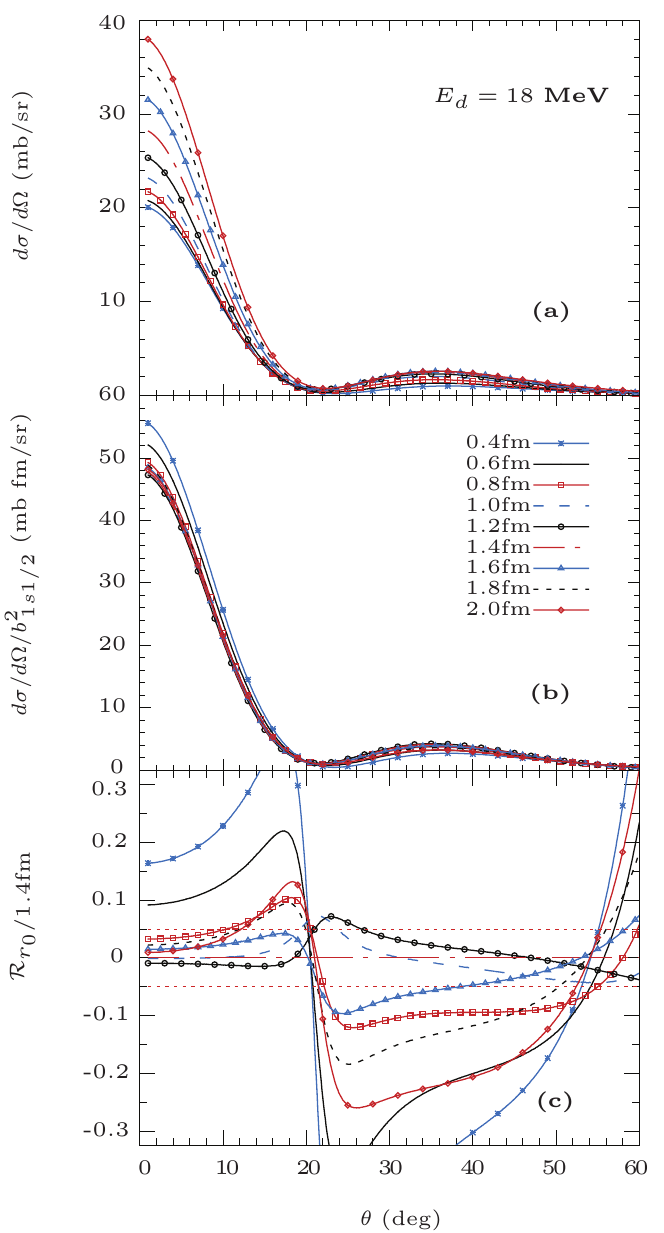}
\caption{Same as Fig.~\ref{f4} for $E_d=18$~MeV.}
\label{f5}
\end{figure}

We repeat our calculations and analysis at the other energies at which data were taken \cite{schmitt,schmittc}.
The results obtained at $E_d=18$~MeV are presented in \Fig{f5}.
As at 21.4~MeV, the reaction is peripheral at forward angles.
However, the region of peripherality is enlarged up to $\theta<10^\circ$ and even though the short-range potentials still lead to significant ratios ${\cal R}_{r_0/1.4{\rm fm}}$, they move closer to the 5\% acceptance band.
It seems therefore that transfer reactions measured at lower beam energy are more peripheral.

\begin{figure}
\centering
\includegraphics[width=\linewidth]{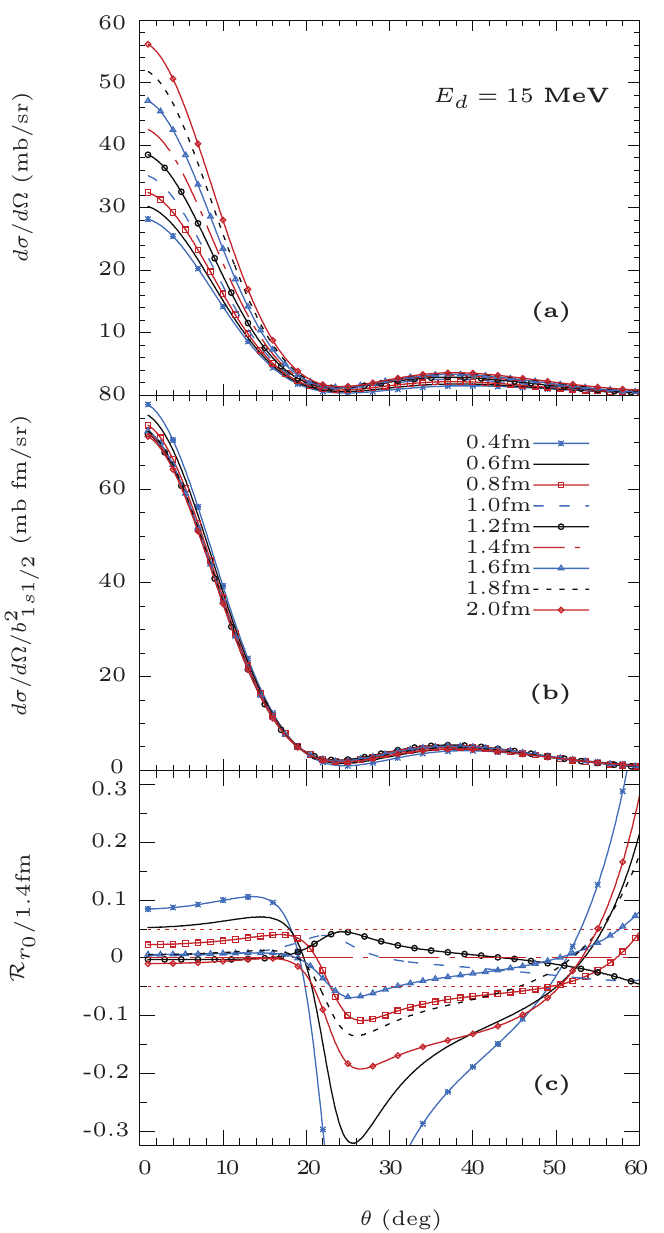}
\caption{Same as Fig.~\ref{f4} for $E_d=15$~MeV.}
\label{f6}
\end{figure}

\begin{figure}
\centering
\includegraphics[width=\linewidth]{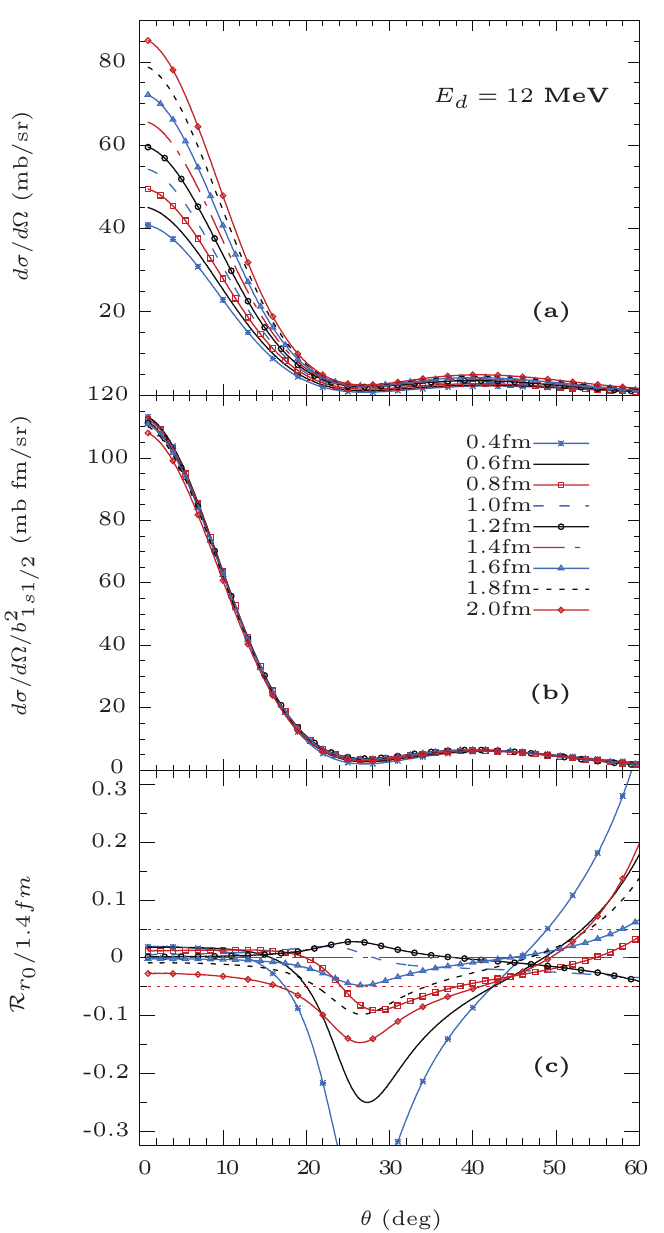}
\caption{Same as Fig.~\ref{f4} for $E_d=12$~MeV.}
\label{f7}
\end{figure}

Moving down in energy confirms this trend.
At $E_d=15$~MeV (\Fig{f6}), the peripherality angular range goes up to $20^\circ$ and the results obtained with the narrow potentials are now within a mere 10\% of the more regular widths.
At even lower energy ($E_d=12$~MeV, \Fig{f7}), the peripherality at forward angle is even clearer.
This can already be seen in \Fig{f7}(b), and the panel (c) confirms that all potentials, even the most narrow ones, fall into the peripherality acceptance band for $\theta<20^\circ$.
We therefore conclude that, first, the peripheral area of this transfer reaction is always found at forward angles, and second that when the incident energy decreases, the reaction exhibits a more pronounced peripheral character.

\subsubsection{\label{ANC}Extraction of the ANC of the $^{11}$Be g.s.}

Now that we know in which conditions the reaction is peripheral (low $E_d$ and forward angles), we extract an ANC by scaling our calculations to the data of Schmitt \etal\ in these exact conditions.
For each beam energy, and each potential width $r_0$, we thus infer an ANC $C_{n_{r'}lj}^{(r_0)}$ from a $\chi^2$ analysis 
\begin{equation}
\chi_{(r_0)}^2=\sum_{i'}\frac{\left[ \left(\frac{C_{n_{r'}lj}^{(r_0)}}{b_{n_{r'}lj}^{(r_0)}}\right)^2 \left.\frac{d\sigma_{\rm th}^{(r_0)}}{d\Omega}\right|_{i'}-\left.\frac{d\sigma_{\rm exp}}{d\Omega}\right|_{i'}\right] ^2}{\left(\left.\delta_{\rm exp}\right|_{i'}\right)^2}
\label{e8}
\end{equation}
where $\left.\delta_{\rm exp}\right|_{i'}$ is the experimental uncertainty at angle $\theta_{i'}$ and the sum is limited to the sole data points $i'$ which lie within the peripheral regions defined in the previous section, viz. within the 5\% acceptance band.

\begin{figure}
\centering
\includegraphics[width=\linewidth]{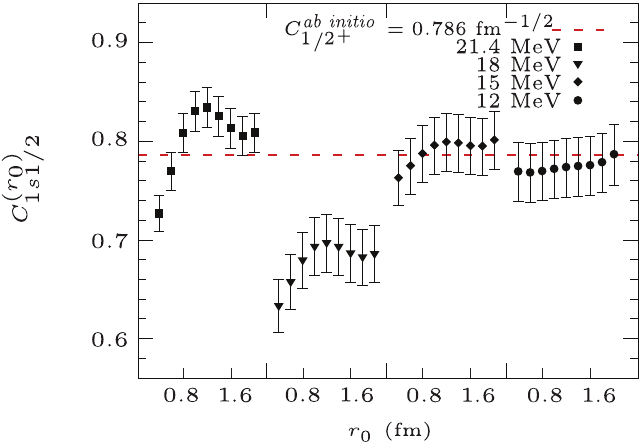}
\caption{ANCs extracted for the ground state of $^{11}$Be by minimizing the $\chi^2$ \eq{e8} for each beam energy and each potential of \tbl{t1}.
The \ai\ result ($C^{\textit{ab initio}}_{1/2^+}=0.786$~fm$^{-1/2}$) is displayed for comparison by the dashed line.}
   	 \label{f8}
\end{figure}

The ANCs $C_{1s1/2}^{(r_0)}$ obtained by minimizing the sum in \Eq{e8} are shown in Fig.~\ref{f8} as a function of the potential width $r_0$ (from $r_0=0.4$~fm on the left to $r_0=2.0$~fm on the right) and are grouped according to the beam energy: $E_d=21.4$~MeV (squares), 18~MeV (triangles), 15~MeV (diamonds), and 12~MeV (circles).
The error bars correspond to the uncertainty in the $\chi^2$ minimization.

The extraction of these ANCs is more reliable at low energy: the dependence on $r_0$ vanishes for the lowest beam energies.
At $E_d=21.4$~MeV, even if one except the results obtained with the shortest widths $r_0$ (first two points), we observe a significant dependence on the potential geometry.
This confirms that, at this energy, even when selecting the data at forward angles, the reaction is not completely independent of the internal part of the radial wave function (see \Fig{f4}).
There is a problem with the results at $E_d=18$~MeV, which are always smaller than at the other beam energies.
This has already been seen in Schmitt \emph{et al.}'s analysis \cite{schmitt,schmittc}.
The reason for that remains unclear.
However, here too, the dependence of the extracted ANC on $r_0$ cannot be neglected, and hence that reaction cannot be considered as purely peripheral.

As already seen above, the best results are obtained at $E_d=15$ and 12~MeV.
Especially in the latter case, the ANC is nearly independent on the geometry of the potential, which gives us confidence that the value hence inferred is close to the real one.

To infer the actual ANC from the Oak-Ridge data, we thus focus on the two lowest beam energies and select only the calculations that fall within the confidence band of 5\% defined in the previous section, which means that we consider all potentials at $E_d=12$~MeV and the potentials with $r_0\ge0.8$~fm at $E_d=15$~MeV.
We hence obtain an average of $C_{1s1/2}=0.785\pm 0.03$~fm$^{-1/2}$.
This value is close to that found by Belyaeva \emph{et al.} \cite{belyaeva} with a coupled-reaction channel model of the reaction.
More interestingly, it is in excellent agreement with the result obtained by Calci \etal\ within their NCSMC calculation of $^{11}$Be structure \cite{abinitio}: $C^{\textit{ab initio}}_{1/2^+}=0.786$~fm$^{-1/2}$ (dashed line in \Fig{f8}). 

\begin{figure}
\centering
\includegraphics[width=\linewidth]{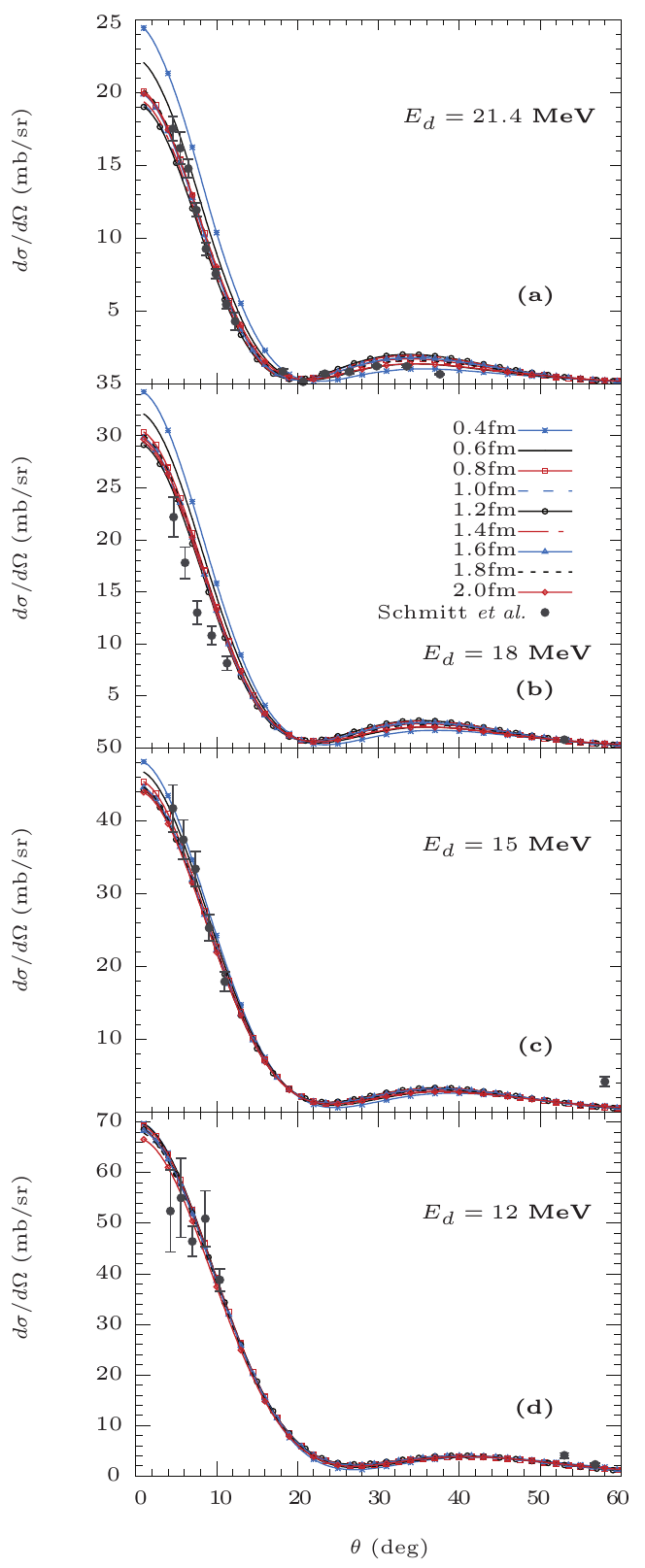}
\caption{The angular distribution for $^{10}$Be$(d,p)^{11}$Be at all experimental energies after scaling to the ANC obtained by the $\chi^2$ minimization $C_{1s1/2}=0.785$~fm$^{-1/2}$.}
   	 \label{f9}
\end{figure}

To estimate the accuracy of the inferred ANC, we plot in \Fig{f9} the results of our calculations scaled to this value, viz. $\left(\frac{C_{1s1/2}}{b_{1s1/2}^{(r_0)}}\right)^2 \frac{d\sigma_{\rm th}^{(r_0)}}{d\Omega}$.
The agreement with the data improves at lower energy, which confirms the method introduced here.
Since this analysis relies a lot on the accuracy of the experimental data, it would be helpful to conduct such experiments focusing on the low energies and forward angles to obtain a more precise ANC.

\subsection{\label{es}Transfer to the $^{11}$Be e.s.}

\begin{figure}[h]
\centering 
\includegraphics[width=\linewidth]{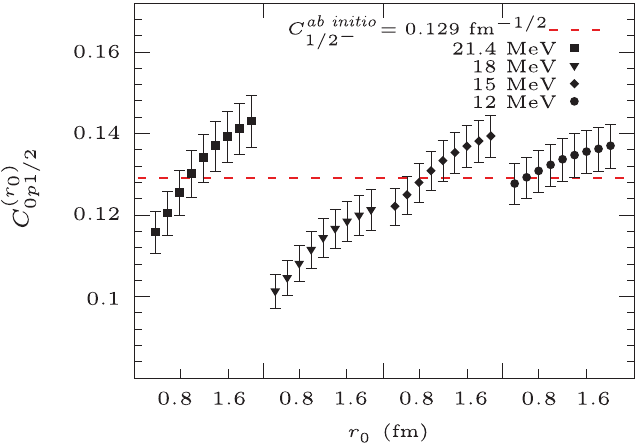}
\caption{ANCs extracted for the $^{11}$Be e.s.\ by minimizing the $\chi^2$ \eq{e8} for each beam energy and each potential of \tbl{t1}.
The \ai\ result ($C^{\textit{ab initio}}_{1/2^-}=0.129$~fm$^{-1/2}$) is displayed for comparison.}
\label{f10}
\end{figure}

We next apply the same method to the data of Schmitt \etal\ on the $^{11}$Be $\half^-$ e.s.
Our results are summarized in \Fig{f10}.
In this case, we observe a much stronger dependence of the results on the potential geometry, and even if it flattens at lowest beam energy, it never becomes negligible at $E_d=12$~MeV.
In our analysis, we have observed a much larger spread of the theoretical cross sections than for the ground state.
This is most likely due to the $p$-wave dominant structure of this state, which, with a non-vanishing centrifugal barrier, forces a large fraction of the wave function to be in the interior of the nucleus, hence leading to transfer reactions that are no longer purely peripheral at these energies.

To infer an ANC from the existing data, we hence focus solely on the set of data at the lowest energy ($E_d=12$~MeV).
As for the g.s.\ we consider only the calculations which fall within the 5\% acceptance band, which excludes the potentials with a width $r_0\le0.8$~fm.
From this analysis of the data, we obtain an averaged $C_{0p1/2}=0.135\pm0.005$~fm$^{-1/2}$.
This value is also comparable to that obtained in \Ref{belyaeva} and is close to the \ai\ value of Calci \etal\ $C^{\textit{ab initio}}_{1/2^-}=0.129$~fm$^{-1/2}$ \cite{abinitio}.
To improve the accuracy of the method, one would need transfer data measured at even lower beam energy.
Extrapolating the tendency observed in \Fig{f10}, it seems that at an energy $E_d<10$~MeV, the reaction will become purely peripheral, leading to a dependence on $r_0$ sufficiently negligible to extract a more reliable ANC.

\subsection{\label{sensitivity}The sensitivity to the optical potential choice}
\begin{figure}
\centering
\includegraphics[width=\linewidth]{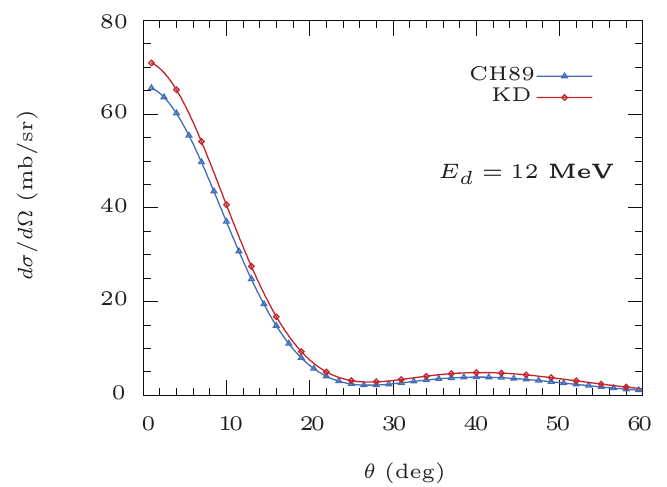}
\caption{Influence of the nucleon-nucleus optical potential on the transfer cross section for $^{10}$Be$(d,p)^{11}$Be at $E_d=12$~MeV.
The Gaussian $^{10}$Be-$n$ potential is chosen with a width $r_0=1.4$~fm in both cases.}
\label{f11}
\end{figure}

All the calculations presented in this work have been obtained using the Chapel-Hill (CH89) global nucleon-nucleus optical potential \cite{ch89}.
However, other choices are possible.
To estimate the sensitivity of our calculations to this potential choice, we repeat our calculations with the Koning-Delaroche potential (KD) \cite{kd}.
This analysis is illustrated in \Fig{f11} for the transfer reaction $^{10}$Be$(d,p)^{11}$Be at $E_d=12$~MeV.
In both cases, we use the Gaussian $^{10}$Be-$n$ potential with a width $r_0=1.4$~fm.

As already observed in \Ref{schmitt,schmittc,MMT14}, we observe that the KD potential leads to a larger cross section compared to the CH89 one.
Besides this change in magnitude of the cross section, the choice of optical potential does not affect the method.
Since the cross sections calculated with the KD potential lead systematically to larger cross sections than those with CH89, we obtain a smaller ANC $C_{1s1/2}^{\rm KD}=0.755\pm0.03$~fm$^{-1/2}$, still in agreement with the \ai\ prediction.

\section{\label{conclusion}Conclusion}
Transfer reactions provide an efficient tool to study the single-particle structure of nuclei away from stability \cite{AN03,thompson,Joh14,euroschool,Pot17,Wim18}.
They are therefore used to study halo structures, like in $^{11}$Be.
In a recent experiment, Schmitt \etal\ have measured the  $^{10}$Be$(d,p)^{11}$Be transfer reaction at $E_d=21.4$, 18, 15 and 12~MeV \cite{schmitt,schmittc}.

We have reanalyzed these data within the ADWA model of transfer \cite{tandy}, using a Halo-EFT description of $^{11}$Be at leading order \cite{BHK02,HJP17}.
This enables us to precisely study the sensitivity of the cross sections to the short-range physics of the $^{10}$Be-$n$ wave function of both the g.s.\ and e.s.\ of $^{11}$Be.
Accordingly, we have been able to define the experimental conditions under which the reaction can be considered as peripheral, and hence from which a reliable ANC can be extracted \cite{BDB77,Gag99,MN05,pang,INY07,MMT14,belyaeva}.

For the $\half^+$ g.s.\ of $^{11}$Be, selecting the data at low energy ($E_d\le15$~MeV) and forward angles ($\theta<20^\circ$) seems enough.
Transfer reactions towards the $\half^{-}$ e.s.\ require a much lower energy to be strictly peripheral, probably because of the existence of the centrifugal barrier in this $p$-wave dominated bound state.
The ideal experimental conditions would actually require $E_d<10$~MeV.

From the comparison between our calculations and the experimental data selected in these conditions of peripherality, we obtain $C_{1s1/2}=0.785\pm0.03$~fm$^{-1/2}$ in the g.s.\ and  $C_{0p1/2}=0.135\pm0.005$~fm$^{-1/2}$ in the e.s.
Both are in excellent agreement with the \ai\ predictions of Calci \etal\ ($C^{\textit{ab initio}}_{1/2^+}=0.786$~fm$^{-1/2}$ and $C^{\textit{ab initio}}_{1/2^-}=0.129$~fm$^{-1/2}$) \cite{abinitio}.
This, adding to the fact that the same value of the g.s.\ ANC leads to excellent agreements with breakup measurements of $^{11}$Be \cite{varna17,CPH18,MC18}, confirm the accuracy of Calci \etal's predictions.

In conclusion, this work suggests a new, systematic and reliable way to extract from transfer measurements the ANC of loosely bound nuclei, e.g. exhibiting a halo.
Our study indicates that investigating transfer reactions at low beam energies and forward angles  ensures the reaction to be peripheral, and is hence the best way to obtain a reliable ANC from experimental data.
This strong constraint on the asymptotics of these nuclei will help investigate the short-range physics of these nuclei as suggested in Refs.~\cite{MN05,pang}.
In a near future, we plan to apply this method to other systems, like $^{15}$C, for which there exist precise data measured at low energy \cite{Gos75,MMT14}.
It would also be interesting to see if this idea can be extended to resonances, like the $\fial^+$ state in $^{11}$Be.

\begin{acknowledgments}
We thank A.~M.~Moro and D.~Y.~Pang for their support in doing the calculation.
This project has received funding from the European Union’s Horizon 2020 research and innovation programme under grant agreement No 654002, the PRISMA (Precision Physics, Fundamental Interactions and Structure of Matter) Cluster of Excellence, and
the Deutsche Forschungsgemeinschaft through the Collaborative Research Center 1044.
J.~Y.\ is supported by the China Scholarship Council (CSC).
P.~C.\ is supported by the Federal State of Rhineland-Palatinate.
This text presents research results of the Belgian Research Initiative on eXotic nuclei (BriX), Program No. P7/12 on interuniversity attraction poles of the Belgian Federal Science Policy Office.
\end{acknowledgments}

\bibliography{transfer}

\end{document}